\documentclass{icrc2009}

\usepackage{graphicx}   
\usepackage[caption=false]{caption}    
\usepackage[font=footnotesize]{subfig} 
\usepackage{fixltx2e}
\usepackage{url}

\newcommand{\shorttitle}[1]%
{\markboth{Proceedings of the 31\MakeLowercase{$^{st}$} ICRC, {\L}\'{o}d\'{z} 2009}{#1} }
\newcommand{\etal}{\MakeLowercase{\textit{et al. }}} 

\begin{document}
\title{Post-launch performance of the Fermi Large Area Telescope}

\author{\IEEEauthorblockN{Riccardo Rando\IEEEauthorrefmark{1},
			  on behalf of the Fermi LAT collaboration}
                            
\IEEEauthorblockA{\IEEEauthorrefmark{1}University of Padova and INFN Padova}

}

\shorttitle{Rando \etal Fermi LAT on-orbit performance}
\maketitle

\begin{abstract}

 The Large Area Telescope (LAT) on-board the {\it Fermi} Gamma-ray Space Telescope started nominal operations on August 13, 2008, after about 60 days of instrument checkout and commissioning and is currently performing an all-sky gamma-ray survey from 30 MeV to above 300 GeV with unprecedented sensitivity and angular resolution.

The LAT pre-launch response was tuned using Monte Carlo simulations and test beam data from a campaign necessarily limited in scope. This suggested a conservative approach in dealing with systematics that affect the reconstruction analysis of the first months of data taking. The first major update of the instrument performance based on flight data is now being completed. Not only are the LAT calibrations now based on flight data, but also the ground event reconstruction has been updated to accommodate on-orbit calibrations, and response was carefully verified using real data from celestial sources. In this contribution we describe the current best knowledge of the instrument, and our plans towards releasing public response functions to support data release in year 2.

  \end{abstract}

\begin{IEEEkeywords}
instruments, instrument performance, {\em Fermi} Gamma-ray Space Telescope
\end{IEEEkeywords}
 
\section{Introduction}
The Large Area Telescope (LAT) is an imaging high-energy $\gamma$-ray telescope, the primary instrument on-board the {\it Fermi} Gamma-ray Space Telescope. It covers the energy range from below 30~MeV to above 300~GeV. For more details on the LAT see \cite{latpaper}.

LAT response was investigated prior to launch using data collected during a test-beam campaign performed in 2006; see e.g. \cite{tbeampaper}. Due to experimental constraints, but most importantly due to the use of a {\it calibration unit} different in geometry with respect to the full-scale LAT, this evaluation of the reconstruction routines was somewhat limited in scope, not allowing a direct check of the LAT event analysis. The test-beam campaing allowed us to validate the Monte Carlo model of the LAT and to fine-tune several details of the LAT reconstruction and event analysis. LAT calibration was first performed on-ground using muon data, while a further more detailed calibration was scheduled to be carried out in orbit.

The calibration effort started immediately after {\it Fermi} launch on 2008, June $11^{\rm th}$ and the first results are described in great detail in \cite{calibpaper}. The LAT performance was fine-tuned by optimising internal delays and synchronizations, alignment constants and absolute timing. In addition, the determination of detector thresholds, gains and noise was performed and in general a robust understanding of the systems' response was obtained.

\section{LAT event analysis}
LAT response is defined by the reconstruction and data analysis procedure. We document here the current status, based on the so-called {\it P6 analysis}. This was already described in \cite{latpaper}, here we summarize briefly the key points.

After triggering and on-board filtering, accepted candidate photons are downlinked to Earth, where they undergo the full event reconstruction and data analysis. In this stage several different estimates for the direction and energy of the primary are calculated. Once the variables describing the event are all made available, the so-called event analysis is performed. First the best determination of the event energy is decided, choosing from the available methods using automated classification algorithms. Then, the best estimate of the incoming photon direction is similarly chosen. For both figures a corresponding {\it probability} is calculated, expressing the degree of confidence that the chosen values do not lie far from the core of the corresponding distribution. After energy and direction are selected, an additional background rejection stage is applied, improving the on-orbit filtering. To do this, information from all LAT subsystems is examined in detail and several figures-of-merit are evaluated using automated data-mining techniques. All these automated algorithms are trained on detailed Monte Carlo simulations of the behavior of $\gamma$ and background particles hitting the LAT.

Summarising, as a result of the on-ground reconstruction analysis we determined the photon energy and direction for each event, with corresponding confidence levels, and several estimates of the probability the event describes, after all, a $\gamma$-ray and not a background particle. We can now define {\it event classes} in terms of cuts on the high-level parameters we have defined, obtaining a purer $\gamma$ dataset with enhanced spatial and energy resolution as the cuts become harder and harder. As the improvement of the performance is obtained by removing events from the datasets enforcing harsher and harsher cuts, the obvious trade-off is between efficiency, purity and resolution.

\section{Instrument response functions}
The LAT response is expressed by means of {\it instrument response functions} (IRFs). Canonically the detector response is factored into three terms: efficiency in terms of the detector {\it effective area}, and resolutions as {\it point spread function} and {\it energy dispersion}; see e.g. \cite{xIRFpap}. Components of the IRFs are usually a tabular representation of the corresponding figures of merit in terms of the photon true energy and direction in the detector system of reference.

To evaluate the LAT response a dedicated Monte Carlo simulation is performed. A huge amount of $\gamma$ events is simulated, in order to cover with good statistics all possible photon inclinations and energies. All detector volumes and all physics interaction must be simulated, so this is actually a considerable effort in terms of machine time. Background events are not simulated: a separate simulation of the background events, where fluxes are downscaled by a sensible factor to make it runnable in some acceptable time interval, is usually performed to allow the estimation of the contamination of any events class.

Pre-launch performance estimates and related IRF set, henceforth called ``P6\_V1'', are documented in \cite{latpaper}.

\section{On orbit performance}
The outcome of the on-orbit calibration campaign described in \cite{calibpaper} was implemented in the Monte Carlo simulations used for IRF generation. 

While examining downlinked events it become clear that some unexpected interactions between background and $\gamma$ events happened, due to the time evolution of the energy deposition in the detector, the timing of the electronics and of the trigger system, and the details of the reconstruction analysis. This was not observed in Monte Carlo simulation as each event is generated independently and interactions between subsequent events is not possible.

\begin{figure}[!t]
  \centering
  \includegraphics[width=8cm]{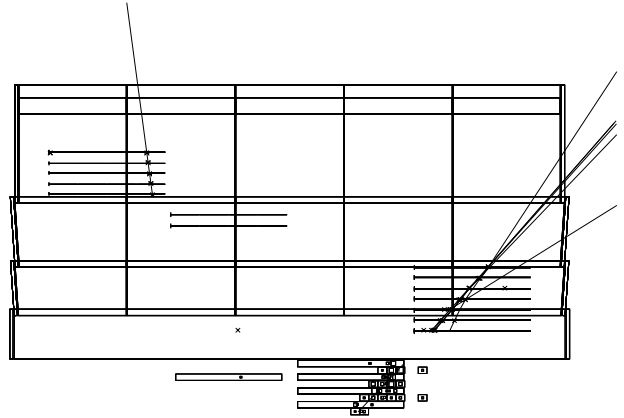}
  \caption{Example of a ghost event in the LAT. On the right, a candidate gamma event, on the left a {\it ghost} track in the tracker, resulting from an accidental time coincidence of a background event; in the tracker and calorimeter only active volumes with some energy deposition are shown, candidate tracks are also drawn.}
  \label{ghost_fig}
 \end{figure}

Several different instances of so-called {\it ghost events} were observed in real data. To make an example, let us consider a background event releasing energy in the detector active volumes. Most background events are easily recognisable as such, so we can assume that a trigger request is not issued and the LAT remains in an active state, waiting for a $\gamma$ event. A ghost may occur if a photon strikes the LAT while the energy released by the background particle is still being collected from sensitive volumes: if a $\gamma$ event triggers the data acquisition and LAT channels are latched and read, digitized signals caused by both the photon and the background hit are obtained and transmitted to Earth. When looking at the downlinked event we see the $\gamma$ event, plus artifacts due to the ghost background hit. See Fig.~\ref{ghost_fig} for an example of a ghost event.

A certain amount of perfectly legitimate $\gamma$-ray events are therefore muddled by spurious signals. When the reconstruction analysis routines, trained on data samples unaffected by ghost hits, process these data, these events may lead to a degradation of the LAT performance. Notably they may end up being discarded by the background rejection stage, resulting in a reduced $\gamma$ efficiency.

\section{On orbit response of the LAT}
Of course the most desirable procedure to circumvent this problem is to redesign the reconstruction analysis routines to correctly treat the events with artifacts. This is currently being investigated and will require some time. The current scenario sees the release of the first ghost-ready reconstruction and event analysis packages take place before the end of 2009.

On a shorter timescale, to allow for an unbiased analysis of celestial sources, a possible solution is to include the effect of ghost hits in the Monte Carlo simulation used for IRF generation, so that the data selection employed for the performance estimate simulates also the inefficiencies affecting science data. This way estimated IRFs are once again a faithful description of the LAT performance. 
To prepare the ghost hits to be added to simulated gamma events the periodic trigger is used. This is one of the trigger primitives active onboard the LAT: a trigger request is automatically issued every 0.5~s, the LAT is then readout and the obtained data downlinked bypassing the onboard filters. Periodic data are then stored in bins of magnetic latitude to correctly describe the effect of a varying rigidity cutoff on the background rates and spectra. The event frames (most are actually empty and only a fraction contains particle hits) are randomly superimposed to Monte Carlo simulated $\gamma$ events by randomly picking a dataframe from the bin corresponding to the magnetic latitude of the LAT along the simulated orbit. Obtained data are then reconstructed and analyzed using the existing P6 event analysis routines, all cuts defining event classes are applied and from the obtained dataset IRFs are evaluated.

This way the ghost-affected events are not recovered in the event reconstruction and analysis, but the performance degradation is accounted for in the IRFs. The spectra that are calculated applying these IRFs to the LAT data are no more affected by a systematic overestimate of the LAT efficiency. 
Of course at this stage spatial and energy resolutions are still not optimal given the degradation caused by ghost tracks in the tracker and ghost depositions in the calorimeter, but the effect is limited as we will see in the following sections. Besides, while the current treatment returns a good description of the LAT acceptance averaged in time, residual effects are documented on short time scales, as acceptance shows a dependence from the background rates along the orbit. These effects should disappear once a full treatment of the ghost is implemented in the event reconstruction and analysis.

\section{Estimated on-orbit response}

Following the procedure described in the previous section a set of IRFs was produced, the first to include on-orbit effects. We call these IRFs ``P6\_V3'' and, following what was done with pre-flight response, we define three event class with increasing tighter requirements on the background rejection efficiency: Transient, Source and Diffuse\footnote{A similar IRF set, including azimuthal dependence for the effective area has also been generated (``P6\_V5''); efficency shows a 4-fold symmetry over the azimuthal angle around the LAT $z$ axis, with a variation of the order of a few percent. All plots presented in this contribution are identical for P6\_V5 and P6\_V3 IRFs.}. For more details see \cite{latpaper}. 

\begin{figure}[!t]
  \centering
  \includegraphics[width=8cm]{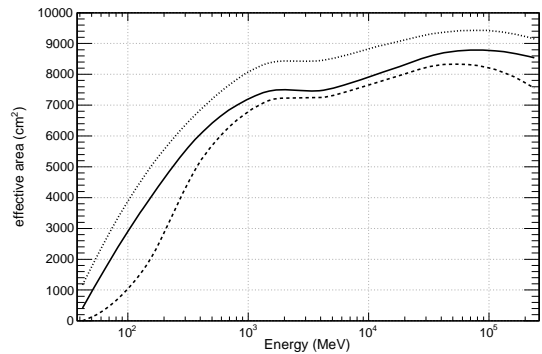}
  \caption{Effective area versus energy at normal incidence for Diffuse (dashed), Source (solid) and Transient (dotted) P6\_V3 event classes.}
  \label{aeff3_fig}
 \end{figure}

\begin{figure}[!t]
  \centering
  \includegraphics[width=8cm]{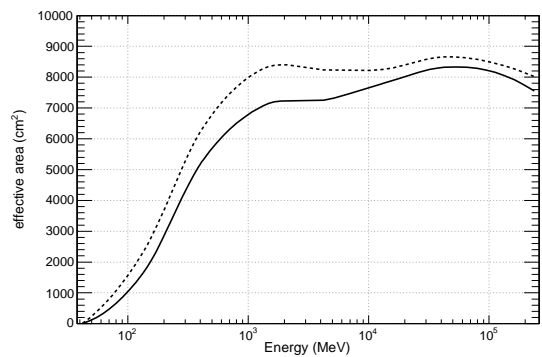}
  \caption{Effective area versus energy at normal incidence for P6\_V1 Diffuse (dashed) and P6\_V3 Diffuse (solid) event classes.}
  \label{aeff_v1v5_fig}
 \end{figure}

In Fig.~\ref{aeff3_fig} we plot the effective area for normally incident photons as a function of energy for the three event classes currently maintained by the LAT collaboration. Effective area is averaged over all azimuthal angles; some smoothing is applied. 
In Fig.~\ref{aeff_v1v5_fig} we plot a direct comparison of normal-incidence effective area for the Diffuse class: the solid curve is the current estimate, the dashed one is P6\_V1. 
Here we observe on purpose the event class with the tighter constraints, so the effect of ghost contamination is very clear. In particular the relative decrease in effective area is sizable at low energy, as a consequence of the spectrum of ghost events and of details of the gamma-selection criteria. It is important to note that the decrease in effective area with respect to pre-flight estimates lies within the level of systematics evaluated for pre-flight performance: the efficiency degradation is estimated to be less than 20\% above 200 MeV. 
We remind the reader once more that this is a snapshot of the current status while significant improvements are being carried on. 

\begin{figure}[!t]
  \centering
  \includegraphics[width=8cm]{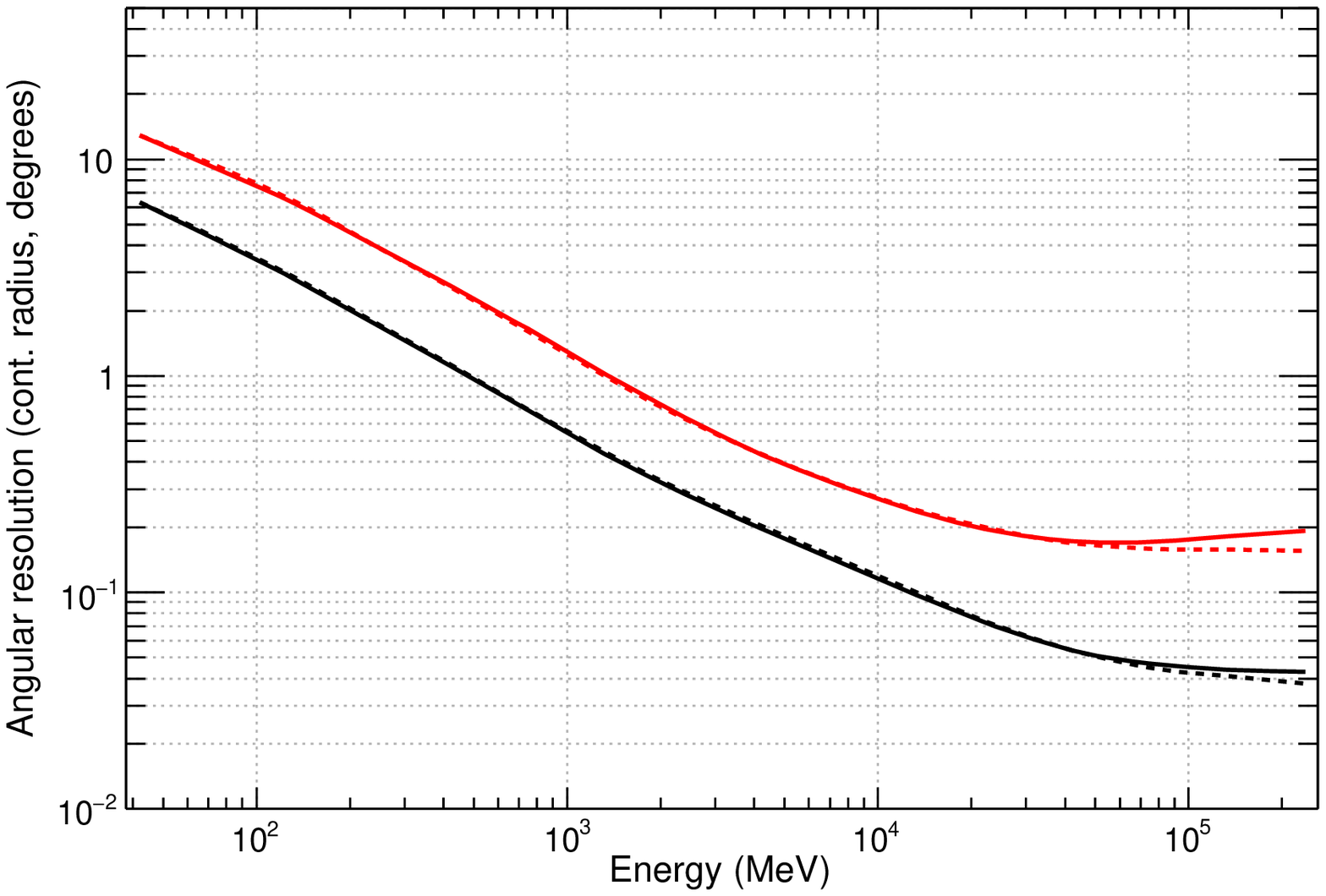}
  \caption{68\% and 95\% PSF containment ratio versus energy at normal incidence for P6\_V1 Diffuse (dashed) and P6\_V3 Diffuse (solid) event classes for conversion in the thin section of the tracker.}
  \label{psf_v1v5_fig}
 \end{figure}

\begin{figure}[!t]
  \centering
  \includegraphics[width=8cm]{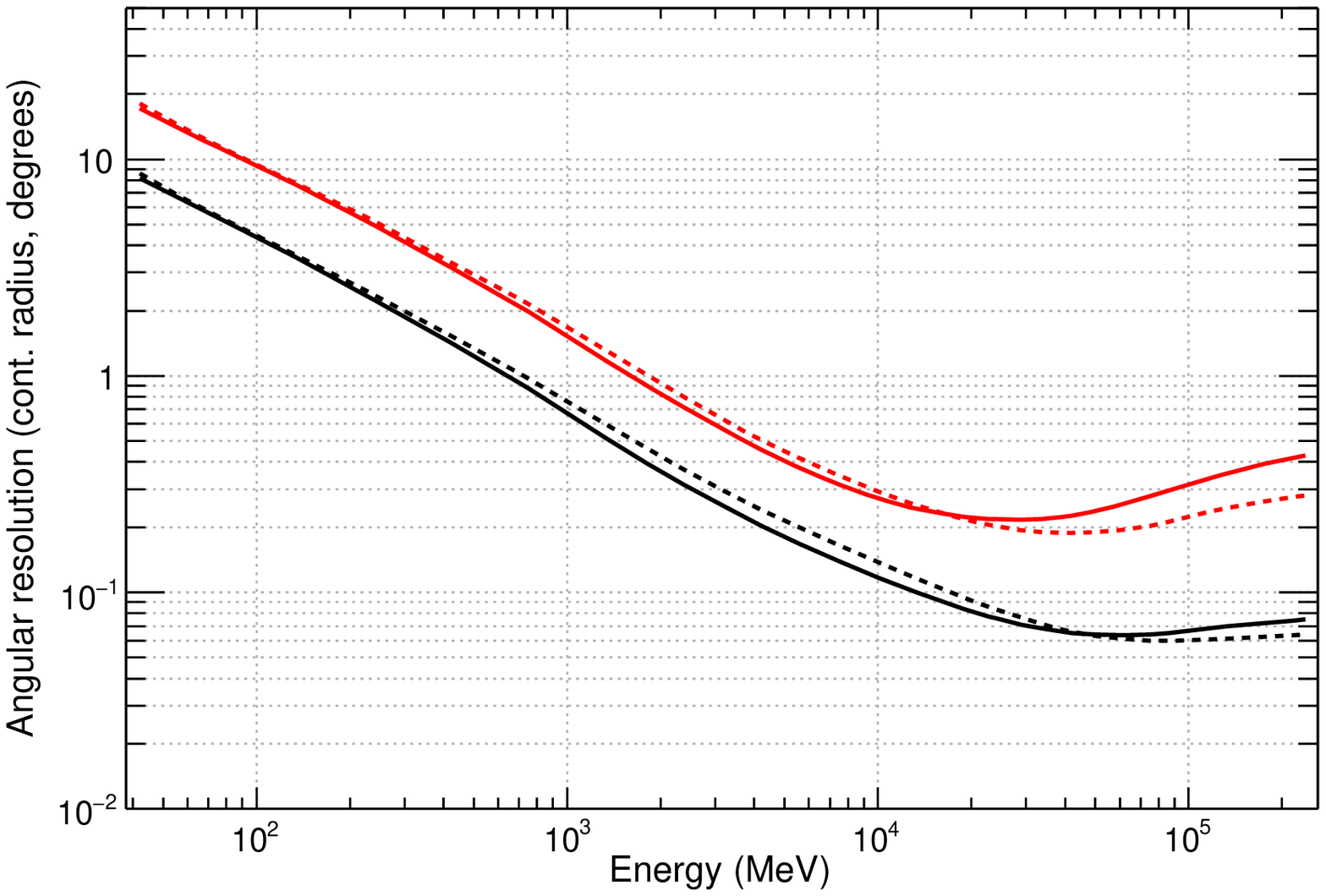}
  \caption{68\% and 95\% PSF containment ratio versus energy at 60 degrees incidence for P6\_V1 Diffuse (dashed) and P6\_V3 Diffuse (solid) event classes for conversion in the thin section of the tracker.}
  \label{psf_v1v560_fig}
 \end{figure}

Both spatial and energy resolution change very little.
In Fig.~\ref{psf_v1v5_fig} (resp. Fig.~\ref{psf_v1v560_fig}) we compare angular resolution as currently estimated (solid line) with P6\_V1 pre-flight estimates. Shown here is the angular resolution versus energy for photons impinging normally (resp. at 60 degrees) for Diffuse class, requiring conversion in the thin layers of the tracker; black curves refer to 68\% containment, red lines to 95\% containment. For 60 degrees incidence at intermediate energies the current IRFs show a clear improvement with respect to pre-flight estimates. We notice that, while the new simulations on which P6\_V3 IRFs are based include several improvements besides the introduction of ghost effects, the effect of ghost tracks could favor the rejection of events that lie in the tails of the distributions, thus causing a small improvement in the angular resolution correlated to the decrease in efficiency.

\begin{figure}[!t]
  \centering
  \includegraphics[width=8cm]{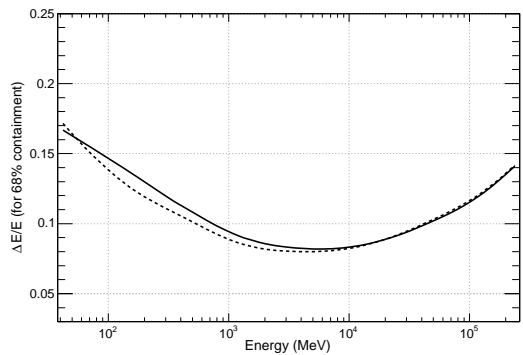}
  \caption{68\% energy resolution versus energy at normal incidence for P6\_V1 Diffuse (dashed) and P6\_V3 Diffuse (solid) event classes.}
  \label{edisp_v1v5_fig}
 \end{figure}

\begin{figure}[!t]
  \centering
  \includegraphics[width=8cm]{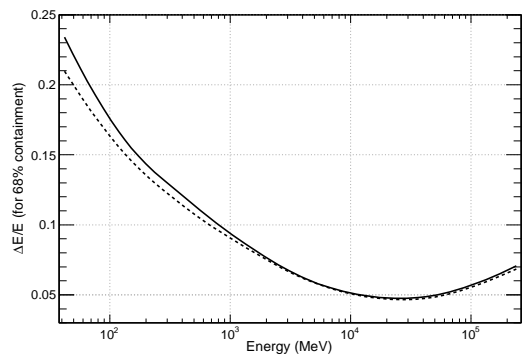}
  \caption{68\% energy resolution versus energy at 60 degrees incidence for P6\_V1 Diffuse (dashed) and P6\_V3 Diffuse (solid) event classes.}
  \label{edisp_v1v560_fig}
 \end{figure}

In Fig.~\ref{edisp_v1v5_fig} (resp. Fig.~\ref{edisp_v1v560_fig}) we compare energy resolution as currently estimated (solid line) with the latest pre-flight estimate (dashed); shown here is the energy resolution versus energy for photons impinging normally (resp. at 60 degrees) for Diffuse class. Some smoothing is applied.

Systematics affecting the LAT efficiency can be estimated by analyzing bright sources; in addition the use of pulsars allow us to select high-purity $\gamma$ samples using the timing information and restricting the selection to the peaks in the pulsar light curve. With this method we currently obtain a conservative estimate for the Diffuse class that puts a 10\% upper limit on flux systematics at 100 MeV, 5\% at 500 MeV and 20\% at 10 GeV.

\section{IRF release plan}

Event analysis routines are currently being updated to into a new scheme (``P7''). This will include a dedicated analysis of background events like electrons, protons and heavy ions; several event classes will be made available for science analysis. Moreover, all automated predictors and classifiers used in the analysis scheme will be retrained using Monte Carlo datasets including ghost affects, so a significant improvement of the performance is expected.

In addition, steps are being taken to manage the presence of ghost-affected events within the reconstruction and event analysis stages. It was successfully demonstrated that ghost tracks in the tracker can be recognised and tagged, while a new clustering scheme in the calorimeter is being investigated. Once the necessary changes in the reconstruction algorithms are complete, specific cuts and bypasses can be added to the analysis scheme to recover the affected events and evaluate the optimal angular and energy resolutions.

``P7'' event analysis will be ready by May 2009. We foresee the first tests on the updated reconstruction algorithms this Fall, and the treatment of ghost effects in the event analysis before the end of 2009. 

\section{Conclusion}
After the early calibration phase the first on-orbit performance estimates for the {\it Fermi} LAT are evaluated. Currently LAT response suffers from some degradation due to residual energy depositions caused by the high background particle rate, rather evident in the case of the effective area, and it is to be considered not yet optimal. This is currently addressed with the release of the IRF set described in this contribution. We have documented the current status of reconstruction and data analysis, and we discussed the future updates being implemented to recover more of the full capability of the LAT. 

\section*{Acknowledgement}
The {\it Fermi} LAT Collaboration acknowledges support from a number of agencies and institutes for both development and the operation of the LAT as well as scientific data analysis. These include NASA and DOE in the United States, CEA/Irfu and IN2P3/CNRS in France, ASI and INFN in Italy, MEXT, KEK, and JAXA in Japan, and the K.~A.~Wallenberg Foundation, the Swedish Research Council and the National Space Board in Sweden. Additional support from INAF in Italy for science analysis during the operations phase is also gratefully acknowledged.

\end{document}